\pdfoutput=1
\documentclass[12pt]{article}
\usepackage{mathrsfs,amsmath,amsfonts,mathtools,amssymb,wasysym,graphicx,subfigure,caption,rotating,comment,booktabs,cite,bm,color,subcaption,authblk}

\usepackage[top=2.0cm,
            bottom=2.0cm,
            left=2.54cm,
            right=2.54cm]{geometry}

\usepackage[hyperindex=true,
          pdfstartview=FitH,
          bookmarksnumbered=true,
          bookmarksopen=true,
          citecolor=blue,
          linkcolor=blue,
          colorlinks=true,
          pdfborder=001,
          unicode]{hyperref}

\allowdisplaybreaks
\DeclareMathOperator{\e}{e}
\newcommand{\D}{{\rm d}}

\newcommand\la{\left\langle}
\newcommand\ra{\right\rangle}
\renewcommand\({\left(}
\renewcommand\){\right)}
\renewcommand\[{\left[}
\renewcommand\]{\right]}
\newcommand\arcsinh{\text{sinh}}

\begin{document}
\title{Adiabatic Elimination in \\Relativistic Stochastic Mechanics}%Multiscale coarse-graining of relativistic stochastic motion and analysis of its implications
\author[1]{Tao Wang \thanks{ {\em email}: \href{mailto:wangtao97@ustc.edu.cn}{wangtao97@ustc.edu.cn}
% \href{taowang@mail.nankai.edu.cn}{taowang@mail.nankai.edu.cn}}
}}
\author[2]{Yu Shi\thanks{ {\em email}: \href{mailto:yu\_shi@ustc.edu.cn}{yu\_shi@ustc.edu.cn}}}
\affil[1]{Department of History of Science and Scientific Archaeology, University of Science and Technology of China, Hefei 230026, China}
%\affil[2]{University of Science and Technology of China, Hefei 230026, China}
\affil[2]{Wilczek Quantum Center, Shanghai Institute for Advanced Studies, University of Science and Technology of China, \protect\\ Shanghai 201315, China}

\date{}
\maketitle

\begin{abstract}
We investigate the adiabatic elimination of fast variables in relativistic stochastic mechanics, which is analyzed by using the equation of motion and the distribution function, with relativistic corrections explicitly derived. A new dimensionless parameter is introduced to characterize the timescale.  The adiabatic elimination is compared with the path integral coarse graining, which is  more general yet computationally demanding. 
\vspace{1em}

\end{abstract}

\section{Introduction}

Diffusion is ubiquitous and is a typical example of irreversibility, which is a  central subject in thermodynamics. Relativity introduces covariance as a basic constraint for fundamental theories and is a basic element in understanding space-time. Both thermodynamics and relativity are referred to as the principle theories by Einstein \cite{einstein1919times}. As a combination of the two,  relativistic diffusion is  a fundamental problem. It has been discussed in a variety of contexts \cite{hakim1965covariant,dudley1966lorentz,debbasch1997relativistic,dunkel2009relativistic,herrmann2010diffusion,smerlak2012diffusion,kremer2014diffusion,dunkel2005theory1,cai2023relativistic1}, with applications in plasma physics, cosmology, and many other fields \cite{he2023heavy,albertini2025stochastic,giardino2024first}. However, this subject appears to be rather unsettled at present, with the coexistence of  various models and approaches out of different motivations. It remains challenging to establish a definitive experimental benchmark to determine which model, if any, is truly accurate. 

As concluded in a review \cite{dunkel2009relativistic}, in order to achieve a covariant formalism, almost all studies adopt one of two strategies: 1. Consider  non-Markovian diffusion processes in spacetime;  2. Construct  relativistically acceptable Markov processes in phase space. The first approach allows one to directly write down a diffusion equation in spacetime satisfying the requirement of relativity, but the underlying mechanical dynamics of the particle remains unknown. The second approach provides the mechanical equation of motion, but only allows indirect access to the behavior of a Brownian particle in spacetime. More recently, as a development of the first approach,  some authors established the diffusion equation in the spacetime by considering jump processes between time slices of spacetime, and then constructing the corresponding equation based on the requirement of covariance and quadratic growth of mean square displacement\cite{smerlak2012diffusion,smerlak2013einstein}. It reconciles the simplicity of Markovian dynamics with the requirement of relativistic covariance. However, the equation of motion of a relativistic Brownian particle is still missing. {There are also studies that characterize relativistic stochasticity based on the relativistic Liouville equation\cite{gao2019relativistic,everitt2025}. However, that approach is more appropriate for, and often focuses on, generalizing the description to the quantum regime. In this paper, we restrict our attention to the classical case.}

{Adiabatic elimination is a powerful method when analyzing systems with variables evolving on different timescales, in which the characteristic parameters separate the dynamics into distinct scales, and these parameters (or their inverses) can often be treated as infinitesimal quantities from the perspective of the slower timescales. Methods that separate fast and slow variables and remove the fast variables from the final effective description are collectively known as the adiabatic elimination of fast variables.} This method has been systematically studied \cite{van1985elimination}, but its use in relativistic settings has received little attention. Relativistic stochastic mechanics provides a natural context of  applying this technique. One reason is that the covariant formulation of relativistic stochastic mechanics is defined on the mass shell bundle, whereas the familiar observable phenomena occur in spacetime. This necessitates a coarse-graining procedure in order to meaningfully connect theoretical predictions with experimental observations. Another reason is the curiosity about how relativistic thermodynamics degenerates to classical equilibrium thermodynamics. This transition cannot be fully captured by the naive `Newtonian limit'; rather, a more detailed analysis requires a systematic treatment of adiabatic elimination within the relativistic stochastic framework. 

The present paper aims to develop the second approach as described above. We try to deduce the diffusion equation in spacetime from the covariant  framework, by using the method of the adiabatic elimination of fast variables. Here we restrict our analysis to (1+1)-dimensional Minkowski spacetime, where the intrinsic geometry is trivial. This significantly simplifies the analysis.

This paper is organized as follows. We review the (1+1)-dimensional relativistic stochastic mechanics in Sec.~\ref{sec2}, with both the equation of motion and the equation of distribution function presented. Secs.~\ref{sec3} and \ref{sec4} form the main body of the paper. In Sec.~\ref{sec3}, the expected equation of motion is derived, while in Sec.~\ref{sec4},  the distribution function is obtained.  In both sections, the role of physical scales is emphasized, and a short discussion on the relevant scale criteria is provided. In Sec.~\ref{sec5},  given the inherent limitations of the adiabatic elimination of the fast variables,  we  turn to a more general approach based on the path integral formalism, which, in principle, can describe the diffusion in spacetime with sufficient precision. Finally, in Sec.~\ref{sec6},  we conclude this article with a summary and discussion of prospects.

\section{(1+1)-Dimensional Relativistic Random Motion}
\label{sec2}

To describe Brownian motion in relativity, there are two natural choices for the evolution parameter, one is the proper time  $\tau$ of the Brownian particle, the other is the proper time of observers $t$. These two descriptions can be transformed into each other by using the reparameterization scheme\cite{cai2023relativistic1}. The covariant Langevin equation LE $_{\tau}$ in Minkowski spacetime $\mathcal{M}$ with Cartesian coordinates takes the form \cite{cai2023relativistic2}
\begin{align}
\D \tilde{x}_{\tau}^{\mu}&=\dfrac{\tilde{p}_{\tau}^{\mu}}{m}\D \tau,\\
\D \tilde{\breve{p}}_{\tau}^{i}&=\[\mathcal{R}^{i}{}_{\frak{a}}\circ_{S}\D\tilde{w}_{\tau}^{\frak{a}}+\mathcal{F}^{i}_{\text{add}}\D\tau\]+\mathcal{K}^{i\nu}U_{\nu}\D \tau,
\end{align}
where $(x^{\mu}, \breve{p}^{i})$ are coordinates on the mass shell bundle $\Gamma_{m}^{+}$. The tilde symbol $\tilde{\ }$ indicates that the corresponding quantity is a random variable because of Brownian motion. $\mathcal{R}^{i}{}_{\frak{a}}$ represents the stochastic amplitude, $\mathcal{F}_{\text{add}}^{i}$ is the additional stochastic force\cite{klimontovich1994nonlinear},
%\footnote{The appearance of this additional term is explained in .}, 
$\mathcal{K}^{i\mu}$ is the friction (or damping) coefficient, and $U^{\mu}$ denotes the proper velocity of the heat reservoir. The term $\D \tilde{w}_{\tau}^{\frak{a}}$ corresponds to an increment of a Wiener process parametrized by the proper time $\tau$. In a $(1+d)$-dimensional spacetime, Greek indices $\mu,\nu,...$ and Latin indices {$a,b,...,h$} run over $0,1,...,d$, while Latin indices $i,j,..$ run over $1,2,...,d$. Among these, only the Latin indices {$a, b,...,h$} are used as abstract indices, whereas all others denote concrete components. The indices $\frak{a},\frak{b},...$ label independent stochastic increments, and their total number depends on the specific systems under consideration. The superscripts of the momentum $p$ are sometimes written using Greek indices and sometimes Latin ones, which may cause confusion. However, the key point is that the Brownian particle satisfies the mass shell condition, so only three components are independent. This implies the relationship between the coordinate basis of mass shell $(\Gamma_{m}^{+})_{x}$ and tangent space $T\mathcal{M}_{x}$
%\cite{cai2023relativistic2}
\begin{align}
	\(\dfrac{\partial}{\partial\breve{p}^{i}}\)^{a}=\(\dfrac{\partial}{\partial p^{i}}\)^{a}-\dfrac{p_{i}}{p_{0}}\(\dfrac{\partial}{\partial p^{0}}\)^{a}.
\end{align}
Throughout all discussions, the Brownian particle is assumed to satisfy the mass shell condition, and its dynamics are formulated within the mass shell bundle using the coordinates $(x^{\mu},\breve{p}^{i})$. The occasional use of the coordinate of tangent bundle $(x^{\mu},p^{\mu})$ is for notational convenience.

The corresponding covariant equation LE$_{t}$ for observers $\mathscr{O}(Z)$,  within which one of the observer's proper time is $t$,  is
\begin{align}
\D \tilde{y}_{t}^{\mu}&=\dfrac{\tilde{k}_{t}^{\mu}}{m}\gamma^{-1}\D t,\\
\D \tilde{\breve{k}}_{t}^{i}&=\[\hat{\mathcal{R}}^{i}{}_{\frak{a}}\circ_{S}\D\tilde{W}^{\frak{a}}_{t}+\hat{\mathcal{F}}^{i}_{\text{add}}\D t\]+\hat{\mathcal{K}}^{i\nu}U_{\nu}\D t,
\end{align}
where $(y^{\mu},\breve{k}^{i})$ is the same coordinates as $(x^{\mu},\breve{p}^{i})$ on the mass shell bundle, and the different notations are used only to highlight the distinction between the associated parameters. $\gamma=-\lambda Z_{\mu}p^{\mu}/m$ is the Lorentz factor, $\lambda=|\nabla t|$ accounts for the misalignment between the worldlines of the observer family $\mathscr{O}(Z)$ and the coordinate system $\{\partial/\partial x^{\mu}\}$. For simplicity, the coordinate system is bound with the observers $\mathscr{O}(Z)$, so that $\lambda=1$. The strategy adopted here is to first derive the equations in the simplest possible setting, with more general cases to be considered subsequently. $\D \tilde{W}_{t}^{\frak{a}}$ is a stochastic increment obtained through reparameterization, and it represents the increment of a Wiener process parameterized by proper time $t$. All the quantities with $\hat{\ }$ are also introduced by reparameterization, and their relationship to the original quantities is characterized by
\begin{align*} 
\hat{\mathcal{R}}^{i}{}_{\frak{a}}=\gamma^{-1/2}\mathcal{R}^{i}{}_{\frak{a}},\qquad \hat{\mathcal{K}}^{i\nu}=\gamma^{-1}\mathcal{K}^{i\nu},\qquad\hat{\mathcal{F}}^{i}_{\text{add}}=\gamma^{-1}\mathcal{F}^{i}_{\text{add}}-\dfrac{\delta^{\frak{ab}}}{2}\mathcal{R}^{i}{}_{\frak{a}}\mathcal{R}^{j}{}_{\frak{b}}(\gamma^{-1/2}\nabla_{j}^{(h)}\gamma^{-1/2}),
\end{align*}
where $\nabla_{j}^{(h)}$ is the covariant derivative on the mass shell $(\Gamma_{m}^{+})_{x}$, the superscript $(h)$ indicates that the covariant derivative is taken with respect to the metric $h_{ij}$ induced on the mass shell. 

In (1+1)-dimensional Minkowski spacetime with Cartesian coordinates, it is convenient to consider the {homogeneous} and isotropic diffusion in mass shell bundle, LE$_{\tau}$ and LE$_{t}$ take the following forms respectively:
\begin{align}
\D \tilde{t}_{\tau}=\frac{E}{m}\D\tau=\gamma\D \tau,\qquad\D \tilde{x}_{\tau}=\frac{\tilde{p}_{\tau}}{m}\D \tau,\qquad 
\D \tilde{p}_{\tau}=\frac{\sqrt{D}}{m}E\circ_{S}\D \tilde{w}_{\tau}-\frac{\kappa E\tilde{p}_{\tau}}{m^{2}}\D\tau,
\label{langtau}
\end{align}
and
\begin{align}
\D \tilde{t}_{t}=\D t,\qquad\D\tilde{y}_{t}=\frac{\tilde{k}_{t}}{E}\D t,\qquad 
\D\tilde{k}_{t}=\sqrt{\frac{DE}{m}}\circ_{S}\D\tilde{W}_{t}+\frac{D\tilde{k}_{t}}{4Em}\D t-\frac{\kappa}{m}\tilde{k}_{t}\D t.
\label{langt}
\end{align}
where $E=\sqrt{m^{2}+p^{2}}=m\gamma$, $D$ is the diffusion coefficient in momentum space, and $\kappa$ is the damping coefficient. These parameters are related by the Einstein relation $D=2\kappa T$, where $T$ is the temperature of the heat reservoir measured by its comoving observer. The observers $\mathscr{O}(Z)$ have been chosen as the heat reservoir comoving one, $\mathscr{O}(U)$. These two equations provide the foundation for both the theoretical analysis and numerical simulation of relativistic Brownian particle dynamics.

In Minkowski spacetime, it is convenient to use rapidity to describe the motion of particles and the transformation between different reference frame. In general, a normalized timelike vector $V^{\mu}=e_{0'}{}^{\mu}$ may be chosen as the timelike direction of a moving frame $\{e_{\alpha'}{}^{\mu}\}$, such that the proper momentum admits the decomposition
\begin{align}
	p^{\alpha'}=(m\cosh\vartheta,m\sinh\vartheta),
\end{align}
where indices $\alpha'$ label components with respect to the moving frame. In the system under consideration, the proper velocity of the heat reservoir $U^{\mu}$, serves as a natural choice for the characteristic velocity $V^{\mu}$. While other choices are in principle possible, we adopt $V^{\mu} = U^{\mu}$ throughout this work. Given the above setting, the LE$_{\tau}$ expressed in terms of rapidity takes the form 
\begin{align}
\D \tilde{t}_{\tau}=\cosh\vartheta\D\tau,\qquad\D \tilde{x}_{\tau}=\sinh\vartheta\D \tau,\qquad 
\D \tilde{\vartheta}_{\tau}=\dfrac{\sqrt{D}}{m}\D \tilde{w}_{\tau}-\dfrac{\kappa}{m}\sinh\vartheta\D\tau,
\label{langtau-theta}
\end{align}
which is a stochastic {differential} equation(SDE) with additive noise. 

While the equation of motion describes individual trajectories of Brownian particles, a comprehensive understanding of relativistic Brownian motion also requires a statistical description in terms of the distribution function. The corresponding equation is the reduced Fokker-Planck equation
\begin{align}
	\mathscr{L}(\varphi)=-\nabla^{(h)}_{i}\mathcal{I}^{i}[\varphi],
\end{align}
where $\mathscr{L}=\frac{p^{\mu}}{m}\frac{\partial}{\partial x^{\mu}}-\frac{1}{m}\varGamma^{\mu}_{\alpha\beta}p^{\alpha}p^{\beta}\frac{\partial}{\partial p^{\mu}}$ is the Liouville vector and
\begin{align}
	\mathcal{I}^{i}[\varphi]=\mathcal{K}^{i\nu}U_{\nu}\varphi-\dfrac{\mathcal{D}^{ij}}{2}\nabla^{(h)}_{j}\varphi
\end{align}
denotes the heat flux vector representing the energy exchange between the Brownian particle and the heat reservoir. For {homogeneous} and isotropic diffusion in phase space, we have
\begin{align} \mathcal{I}^{i}[\varphi]=\kappa\Delta^{i\nu}(p)U_{\nu}\varphi-\dfrac{D}{2}\Delta^{ij}(p)\nabla^{(h)}_{j}\varphi.
\end{align}
More specifically, within Minkowski spacetime using Cartesian coordinates, the reduced Fokker-Planck equation can be written as
\begin{align}
	\dfrac{p^{\mu}}{m}\dfrac{\partial \varphi}{\partial x^{\mu}}&=-p_{0}\dfrac{\partial}{\partial\breve{p}^{i}}\(\dfrac{1}{p_{0}}\mathcal{I}^{i}[\varphi]\)\notag\\
	&=E\dfrac{\partial}{\partial p}\[\dfrac{1}{E}\(\dfrac{\kappa pE}{m^{2}}\varphi+\dfrac{DE^{2}}{2m^{2}}\dfrac{\partial\varphi}{\partial p}\)\].
\end{align}
Alternatively, when expressing momentum in terms of rapidity, it can be rewritten as
\begin{align}\label{fp-theta}
	\dfrac{p^{\mu}}{m}\dfrac{\partial \varphi}{\partial x^{\mu}}&=\dfrac{\partial}{\partial\vartheta}\(\dfrac{\kappa}{m}\sinh\vartheta \varphi+\dfrac{D}{2m^{2}}\dfrac{\partial\varphi}{\partial\vartheta}\).
\end{align}
In the following discussion, we carry out separate analyses in particle dynamics and in the distribution function.

\section{Elimination of Fast Variables in Equation of Motion }
\label{sec3}

Even in the (1+1)-dimensional case, no exact solution is available for the relativistic Brownian particle. The primary difficulty arises from the complexity of the underlying nonlinear stochastic differential equations. However, within various approximation schemes, one can extract a wealth of useful information regarding its stochastic dynamics. The ultimate goal of a theory is to allow comparison with experimental observations. Therefore, although much of the discussion is initially framed in terms of LE$_{\tau}$, the ultimate focus must lie on LE$_{t}$.

\subsection{Exact solution of relativistic damping motion}
\label{sec3.1}

Since we aim to discuss the elimination of fast variables, essentially a coarse-graining procedure, it is natural to begin with the coarsest level of description, namely by completely neglecting the random force. Then in this case LE$_{\tau}$ becomes damped equation of motion,
\begin{align}
	\D t_{\tau}=\dfrac{E}{m}\D\tau,\qquad\D x_{\tau}=\dfrac{p_{\tau}}{m}\D\tau,\qquad\D p_{\tau}=-\dfrac{\kappa E p_{\tau}}{m^{2}}\D\tau,
\end{align}
and the solution for $p(\tau)$ can be obtained
\begin{align*}
	p(\tau)=\dfrac{2m\sqrt{C}\e^{\kappa\tau/m}}{\e^{2\kappa\tau/m}-C},\qquad(C<1)
\end{align*}
where $C$ is a constant of integration. Combined with the initial condition, the solution takes the form
\begin{align}
	p(\tau)=\dfrac{2m\varsigma}{\e^{2\kappa\tau/m}-\varsigma^{2}}\e^{\kappa\tau/m},
\end{align}
where shorthand
\begin{align}
	\varsigma:=\sqrt{1+2\varepsilon_{0}^{-2}(1-\sqrt{1+\varepsilon_{0}^{2}})}
\end{align}
is introduced for notational convenience with $\varepsilon_{0}:=p(0)/m$ denoting the initial dimensionless momentum. Integrating directly from differential equation $\D x=p/m\D\tau$ yields the result
\begin{align}
	x(\tau)&=x(0)-\dfrac{m}{\kappa}\(\log\dfrac{1-\varsigma}{1+\varsigma}-\log\dfrac{1-\varsigma\e^{-\kappa\tau/m}}{1+\varsigma\e^{-\kappa\tau/m}}\).
\end{align}
Although we have obtained an exact solution in a fully relativistic covariant form, it does not directly provide effective information from the observer's perspective. In the absence of the stochastic force, the reparameterization reduces to a simple substitution $\D\tau=\gamma^{-1}\D t$, leading to the following equation in the observer's frame, 
\begin{align}
	\D y_{t}=\dfrac{k_{t}}{E}\D t,\qquad \D k_{t}=-\dfrac{\kappa k_{t}}{m}\D t.
\end{align}
Compared with Eq.~\eqref{langt}, the second term on the right-hand side of the force equation is omitted, as it originates from the additional stochastic force and therefore does not appear in the damped equation of motion.

The solution to this system of equations is given by
\begin{align}
	k(t)=k(0)\e^{-\kappa t/m},\qquad y(t)=y(0)+\dfrac{m}{\kappa}\(\arcsinh\dfrac{k(0)}{m}-\arcsinh\dfrac{k(0)\e^{-\kappa t/m}}{m}\).
\end{align}
Formally, the momentum of relativistic damped motion may appear identical to that in the Newtonian framework. However, the variable $k(t)$ here denotes the spatial component of the particle's proper momentum vector, and does not directly correspond to the Newtonian spatial momentum. This distinction also explains why the displacement equation involves integrating $k(t)/E$ over time, rather than $k(t)/m$, as in Newtonian mechanics.

\subsection{Regime with equilibrated fast variables}

To investigate the diffusion behavior of a Brownian particle in configuration space, it is common to assume that the momentum relax to equilibrium rapidly, then the momentum part of  LE$_{\tau}$ simplifies as follows, 
\begin{align}\label{stable-p}
	0=-\dfrac{\kappa \tilde{p}_{\tau}}{m}\sqrt{1+\dfrac{\tilde{p}_{\tau}^{2}}{m^{2}}}\D\tau+\sqrt{D}\sqrt{1+\dfrac{\tilde{p}_{\tau}^{2}}{m^{2}}}\circ_{S}\D\tilde{w}_{\tau},\qquad\Longrightarrow\qquad \dfrac{\kappa \tilde{p}_{\tau}}{m}\D\tau=\sqrt{D}\D\tilde{w}_{\tau}.
\end{align}
For SDE with multiplicative noise interpreted in the Stratonovich sense, transforming it into additive SDE by dividing out the stochastic amplitude does not alter the average behavior of the stochastic variables. This is fundamentally due to the fact that the Stratonovich calculus preserves the standard rules of differential calculus, in contrast to the Ito interpretation, which does not\cite{risken1996fokker}.

Inserting Eq.~\eqref{stable-p} into the configuration part of the LE$_{\tau}$ equation leads to
\begin{align}\label{cg-LEtau}
	\D \tilde{t}_{\tau}=\dfrac{\overline{E}}{m}\D \tau,\qquad\D\tilde{x}_{\tau}=\dfrac{\tilde{p}_{\tau}}{m}\D\tau=\dfrac{\sqrt{D}}{\kappa}\D\tilde{w}_{\tau},
\end{align}
where $\overline{E}$ refers to the average energy, evaluated with respect to the momentum distribution after the momentum has reached equilibrium. As discussed in \cite{cai2023relativistic1,cai2023relativistic2}, the distribution function $\Phi_{\tau}(t,x,p)$ defined via LE$_{\tau}$ lacks direct physical interpretation. Nevertheless, its mathematical formulation remains well-defined and permissible. The distribution function $\Phi_{\tau}(t,x,p)$ can be connected with the solution of reduced Fokker-Planck equation by
\begin{align*}
	\varphi(t,x,p)=\int\D\tau \Phi_{\tau}(t,x,p).
\end{align*}
Because of  the linearity of the Fokker-Planck equation, both $\Phi_{\tau}$ and $\varphi$ share the same stationary solution, namely the J{\"u}ttner distribution $\phi(p)\sim \e^{-\beta U_{\mu}p^{\mu}}$. Based on the normalization condition and using the rapidity coordinate introduced previously, the normalization constant is obtained,
\begin{align}\label{phi-p}
	1=\int\dfrac{\D p}{|p_{0}|} \phi(p)=\int\dfrac{\D p}{|p_{0}|} \dfrac{\e^{-\beta U_{\mu}p^{\mu}}}{2\mathsf{J}_{00}(\zeta)}=\int\D \vartheta \dfrac{\e^{-\zeta\cosh\vartheta}}{2\mathsf{J}_{00}(\zeta)},
\end{align}
where $\D p/|p_{0}|$ is the invariant volume element on mass shell, $\zeta=\beta m=m/T$ is the relativistic coldness, and $\mathsf{J}_{nm}(\zeta)$ is defined by
\begin{align}
	\mathsf{J}_{nm}(\zeta):=\int_{0}^{\infty}\D\vartheta \sinh^{n}\vartheta\cosh^{m}\vartheta \e^{-\zeta\cosh\vartheta}.
\end{align}
Once the distribution function is known, computing the average energy $\overline{E}$ after momentum relaxation becomes straightforward,
\begin{align}
	\overline{E}=\int\dfrac{\D p}{|p_{0}|}\dfrac{E\e^{-\beta U_{\mu}p^{\mu}}}{2\mathsf{J}_{00}(\zeta)}=m\dfrac{\mathsf{J}_{01}(\zeta)}{\mathsf{J}_{00}(\zeta)}.
\end{align}
Notice that $\overline{E}$ only depends on the mass of Brownian particle $m$ and the temperature of the heat reservoir $T$, so it is a constant in the regime of stabilized momentum. This means that in this regime, LE$_{\tau}$ is coarse-graining into a trivial diffusion process. Utilizing the relation
\begin{align}
	\D \tilde{t}_{\tau}=\dfrac{\mathsf{J}_{01}(\zeta)}{\mathsf{J}_{00}(\zeta)}\D \tau,
\end{align}
and performing the reparameterization accordingly, the resulted equation of motion for the particle in configuration space from the observer's perspective is given by
\begin{align}\label{cg-LEt}
	\D \tilde{y}_{t}=\dfrac{\sqrt{D}}{\kappa}\sqrt{\dfrac{\mathsf{J}_{00}(\zeta)}{\mathsf{J}_{01}(\zeta)}}\D\tilde{W}_{t}.
\end{align}
From this expression, it is clear that the coarse-graining process breaks the relativistic bound on particle velocity. Although it is not impossible in principle to retain covariance of the equation in spacetime during coarse-graining, no such formulation has yet been successfully achieved in the present framework. Despite these limitations, the relativistic modifications on diffusion remain evident, particularly in the regime of low relativistic coldness $\zeta$. We will present a more intuitive demonstration of this effect through numerical simulations in a later discussion.

\subsection{Overdamped limit of equation of motion}
\label{sec3.3}

The situation considered in the previous subsection, wherein the momentum quickly relaxes to equilibrium, in essence represents the overdamped limit. The validity of this assumption is supported by a timescale analysis, which also naturally leads to a general formulation of the adiabatic elimination method at the level of equations of motion. Physically, this corresponds to a regime in which the observational timescale $\Delta t_{\text{o}}$ is much larger than the characteristic damping timescale. This separation can be expressed in terms of damping parameters as
\begin{align}
	\Delta t_{\text{o}}\gg\dfrac{m}{\kappa}:=\epsilon.
\end{align}
Unlike Sec.~\ref{sec3.1}, where the random force was directly omitted, we now retain its contribution in the analysis. Denoting the characteristic timescale of random force by $\Delta t_{\text{r}}$, the complete timescale hierarchy under consideration reads $\Delta t_{\text{o}}\gg m/\kappa\gg\Delta t_{\text{r}}$\cite{sekimoto2010stochastic}. Within the Newtonian framework, it is natural to perform an expansion of the dynamical variable $p$ in powers of a small parameter $\epsilon$, allowing the equations of motion to be systematically derived order by order. This procedure yields effective equations at different timescales\cite{van1985elimination}. For example, the equation of motion governing the diffusion of a free particle can be directly obtained by expanding the momentum to first order in $\epsilon$. In the relativistic context, a similar procedure can be applied following the same idea. However, compared to the Newtonian case, the ubiquitous presence of the energy $E$, or in other words, the hyperbolic structure imposed by the mass shell constraint, introduces additional complexity into the calculations. The essential features can already be seen by examining the first-order expansion in $\epsilon$,
\begin{align*}
	p&=p^{(0)}+\epsilon p^{(1)},\\
	E&=\sqrt{m^{2}+p^{2}}=\sqrt{m^{2}+\(p^{(0)}+\epsilon p^{(1)}\)^{2}}.
\end{align*}
Extracting each order in $\epsilon$ directly from the expression involving square roots, while not impossible, requires performing an additional Taylor expansion of the square root itself. This procedure, however, breaks the underlying hyperbolic structure dictated by the relativistic mass shell condition that governs particle motion.

Disregarding the violation of relativistic causality introduced by this approach, substituting the ansatz into the equation and collecting terms up to the first two orders in $\epsilon$ yields
\begin{align}
	\epsilon^{-1}:&\quad 0=-\dfrac{\tilde{p}_{\tau}^{(0)}}{\epsilon}\sqrt{1+\(\frac{\tilde{p}_{\tau}^{(0)}}{m}\)^{2}}\D \tau,\\
	\epsilon^{0}:&\quad \D\tilde{p}_{\tau}^{(0)}=\dfrac{\sqrt{D}}{m}\sqrt{m^{2}+\(\tilde{p}_{\tau}^{(0)}\)^{2}}\circ_{S}\D\tilde{w}_{\tau}-\(\sqrt{m^{2}+\(\tilde{p}_{\tau}^{(0)}\)^{2}}+\dfrac{\tilde{p}_{\tau}^{(0)}\tilde{p}_{\tau}^{(1)}}{\sqrt{m^{2}+\(\tilde{p}_{\tau}^{(0)}\)^{2}}}\)\dfrac{\tilde{p}_{\tau}^{(1)}}{m}\D \tau.
\end{align}
The $\epsilon^{-1}$-order equation yields $\tilde{p}_{\tau}^{(0)}=0$, hence, the stochastic variable $\tilde{p}_{\tau}^{(0)}$ is in fact deterministic. Substituting it into the $\epsilon^{0}$-order equation, the equation reduces to
\begin{align}
	0=\sqrt{D}\D\tilde{w}_{\tau}-\tilde{p}_{\tau}^{(1)}\D\tau.
\end{align}
Together with the equation in position space, this yields
\begin{align}
	\D \tilde{x}_{\tau}=\frac{\tilde{p}_{\tau}^{(0)}+\epsilon\tilde{p}_{\tau}^{(1)}}{m}\D\tau=\frac{\epsilon\tilde{p}_{\tau}^{(1)}}{m}\D\tau=\frac{\sqrt{D}}{\kappa}\D\tilde{w}_{\tau},
\end{align}
which coincides with \eqref{cg-LEtau}. Consequently, the subsequent discussion proceeds in full analogy with the previous subsection.

In summary, this subsection serves two purposes: on one hand, it clarifies the relevant timescale underlying the use of the adiabatic elimination method; on the other hand, it presents a more general framework for implementing the method. By expanding dynamical variables in terms of the small parameter $\epsilon$, the approach becomes applicable to a broader class of problems, including those involving external fields. However, some caution is required. The scale parameter $\epsilon=m/\kappa$, adopted here in the same manner as in the non-relativistic context, does not in itself encode any relativistic information or corrections. While this does not significantly affect the order-by-order derivation of the reduced equations, it does leave unresolved the issue of how relativistic effects modify the relevant timescales. A more detailed discussion on this point will be provided later.

\subsection{Numerical simulation}
\label{sec3.4}

One of the key advantages of the Langevin equation is its straightforward implementation in numerical simulations. While the distribution of a Brownian particle at a fixed time in (1+1)-dimensional Minkowski spacetime has been studied in \cite{cai2023relativistic1}, the present work focuses on the time evolution of the Brownian particle's behavior.

Based on LE$_{\tau}$ \eqref{langtau}, LE$_{t}$ \eqref{langt}, and the Newtonian Langevin equation
\begin{align}
	\D \tilde{t}_{t}=\D t,\qquad\D \tilde{y}_{t}=\dfrac{\tilde{k}_{t}}{m}\D t,\qquad \D \tilde{k}_{t}=\sqrt{D}\D \tilde{W}_{t}-\dfrac{\kappa\tilde{k}_{t}}{m}\D t,
\end{align}
with the simplest parameter settings $m=1,\kappa=1,\beta=1$, and $D=2$  according to the Einstein relation,  one can obtain the average behavior of the particle after a period of evolution. To avoid introducing unnecessary notation, the phase-space coordinates $(t,y,k)$ are also used for the Newtonian framework. As emphasized in Sec.~\ref{sec2}, $(x^{\mu},\breve{p}^{i})$ and $(y^{\mu}, \breve{k}^{i})$ represent the same coordinates. The use of different symbols is solely for clarity, to distinguish that their respective equations of motion evolve with respect to different time parameters. In the Newtonian case, the use of $(t, y, k)$ serves to highlight the closer correspondence between the observer's time and Newtonian absolute time. Moreover, when analyzing the distribution function at a fixed time, the coordinates $(t,x,p)$ will be used throughout, as the focus is no longer on the evolution of individual particle trajectories. The simulation time settings are as follows: in each individual simulation, the particle's state was updated with a time step of $0.01$. To examine the long-term evolution, we performed repeated simulations at observation times ranging from $20$ to $2000$ in increments of $20$. For each observation time, $10000$ trajectories were generated to analyze the statistical behavior of the particle.

\begin{figure}[htbp]
\centering
\includegraphics[width=0.5\textwidth]{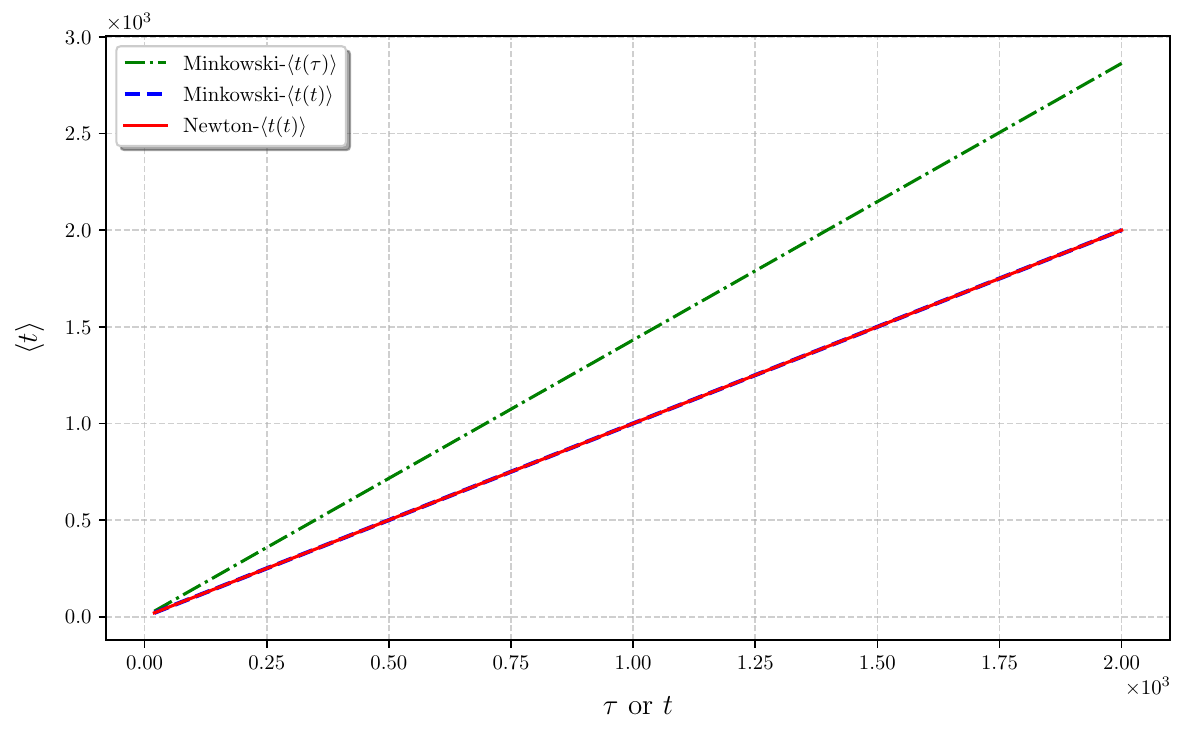}
\caption{Mean coordinate value $t$ as a function of different evolution parameters.}
\label{fig1}
\end{figure}

In the simulations, the first plot containing useful information depicts the mean of coordinate value $t$ as a function of both evolution parameters $\tau$ and $t$(see Fig.~\ref{fig1}). Since the coordinate system is bound to a comoving observer $\mathscr{O}(U)$, the datasets Minkowski-$\la t(t)\ra$ and Newton-$\la t(t)\ra$ naturally align along a straight line of unit slope. The dataset Minkowski-$\la t(\tau)\ra$ also exhibits a linear trend, confirming that under the current parameter settings, the adiabatic elimination method used in the previous subsections is valid. Furthermore, the slope of the curve can be determined analytically and is given by $\mathsf{J}_{01}(1)/\mathsf{J}_{00}(1)\approx 1.42963$.

\begin{figure}[htbp]
  \centering
  \begin{minipage}{0.48\textwidth}
    \centering
    \includegraphics[width=\linewidth]{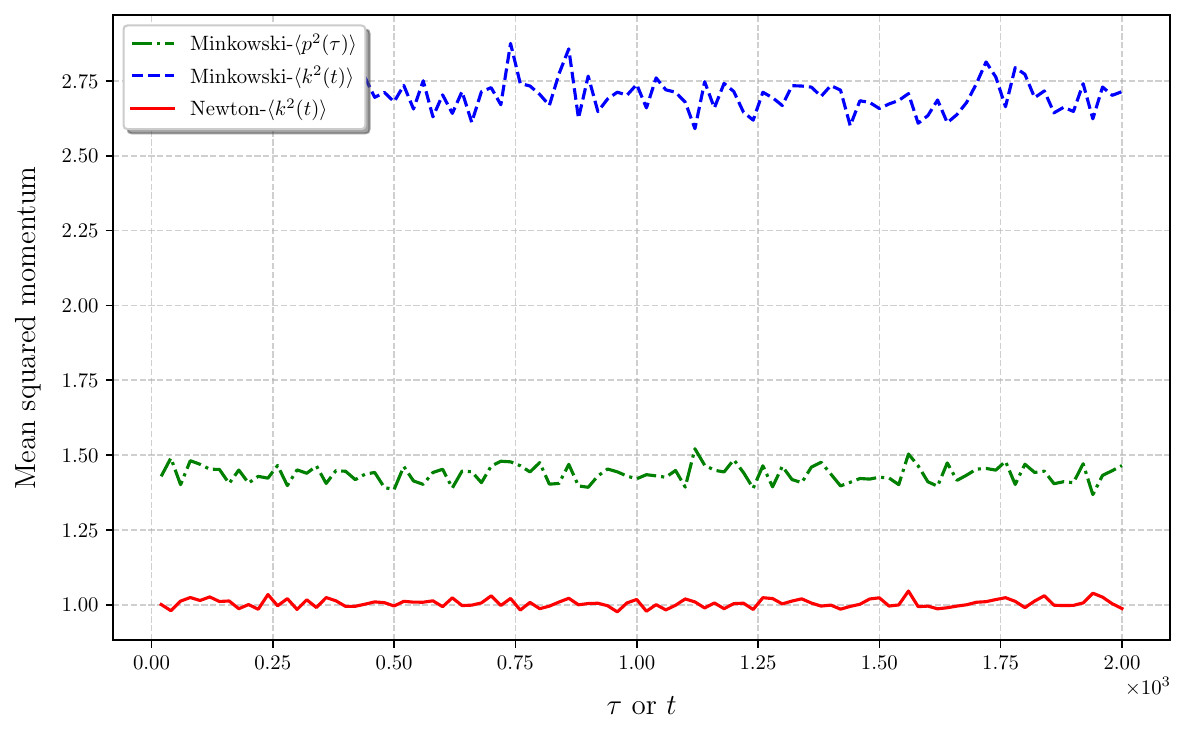}
  \end{minipage}
  \hfill
  \begin{minipage}{0.5\textwidth}
    \centering
    \includegraphics[width=\linewidth]{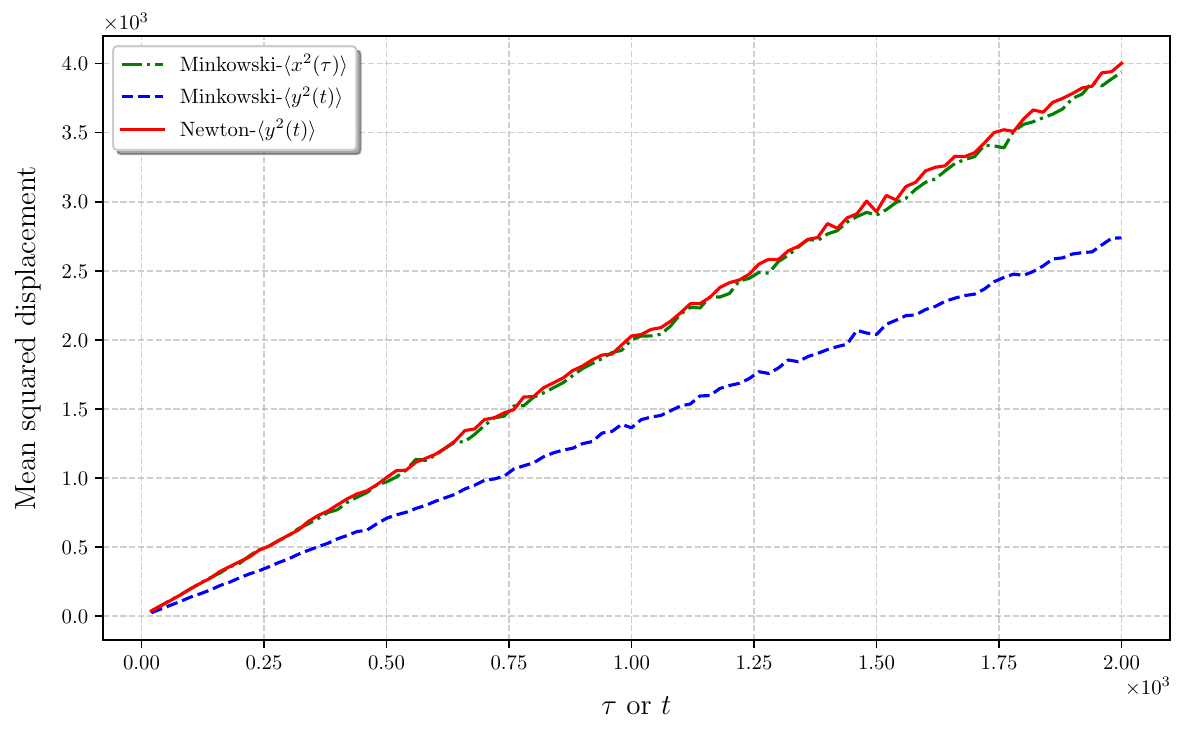} 
  \end{minipage}
  \caption{Mean squared momentum and displacement with respect to the evolution parameter.} 
  \label{fig2}
\end{figure}

Physically, our primary interest is the evolution of the mean squared momentum and mean squared displacement with respect to the evolution parameter, as shown in Fig.~\ref{fig2}. The left subfigure shows that, under the current parameter settings, the momentum rapidly reaches equilibrium in both the Newtonian and relativistic cases. The fact that the mean squared momentum converges to different values may require further explanation. For the Newtonian case, the equilibrium distribution is the Maxwell distribution, from which it follows that the mean squared momentum satisfies $\langle k^{2}(t)\rangle = m/\beta=1$ under the chosen parameters. The distribution $\Phi_{\tau}(t,x,p)$, corresponding to the LE$_{\tau}$, has a stationary solution denoted by $\phi(p)$, as discussed in the previous subsection. Based on $\phi(p)$, the mean squared momentum can be calculated as $\la p^{2}(\tau)\ra=\mathsf{J}_{20}(1)/\mathsf{J}_{00}(1)\approx 1.42963$. Under the adiabatic approximation, the distribution function $f_{t}(x,p)$, corresponding to the LE$_{t}$, can be written as
\begin{align}\label{adia-ft}
	f_{t}(x,p)&\approx\rho(t,x)\mathsf{f}(p).
\end{align}
The stationary distribution appears differently to different observers, a fact that has been investigated in \cite{cai2023relativistic2}. In Minkowski spacetime, we obtain
\begin{align}
	\mathsf{f}(p)&=-\dfrac{1}{Z_{\mathsf{f}}}\dfrac{Z_{\mu}p^{\mu}}{m}\e^{-\beta U_{\mu}p^{\mu}}=\dfrac{\cosh\vartheta}{Z_{\mathsf{f}}}\e^{-\zeta\cosh\vartheta},\\
	Z_{\mathsf{f}}&=-\int\eta_{(\Gamma_{m}^{+})_{x}}\dfrac{Z_{\mu}p^{\mu}}{m}\e^{-\beta U_{\mu}p^{\mu}}=2\mathsf{J}_{01}(\zeta),
\end{align}
where in both equations, the first equality is general, while the second follows from the explicit choice of the observers $\mathscr{O}(Z)$ to be the heat reservoir comoving one. Therefore, the mean squared momentum obtained from the simulation of LE$_{t}$ can be compared with the analytical result $\la k^{2}(t)\ra=\mathsf{J}_{21}(1)/\mathsf{J}_{01}(1)\approx 2.69948$, and the comparison shows good consistency.

Under the present parameter settings, the right subfigure of Fig.~\ref{fig2} shows that the mean squared displacement still exhibits proportionality to the evolution parameters, but the slopes of the curves have shifted, indicating notable differences in the diffusion behavior. The near coincidence of the red and green curves is readily understood, as the coarse-grained LE$_{\tau}$ \eqref{cg-LEtau} and the coarse-grained Newtonian Langevin equation differ only in evolution parameter. The slopes of these two curves are determined by the diffusion coefficient $D/\kappa^{2}=2$. The blue curve can also be explained by using the coarse-grained LE$_{t}$ \eqref{cg-LEt}, with the corresponding slope given by $D/\kappa^{2}\cdot\mathsf{J}_{00}(1)/\mathsf{J}_{01}(1)=1.39897$. Therefore, when the adiabatic elimination method is valid, it can be concluded that the normal diffusion law $\la x^{2}\ra\sim t$ still holds, with relativistic corrections manifested in the proportionality coefficient.

\begin{figure}[htbp]
\centering
\includegraphics[width=0.5\textwidth]{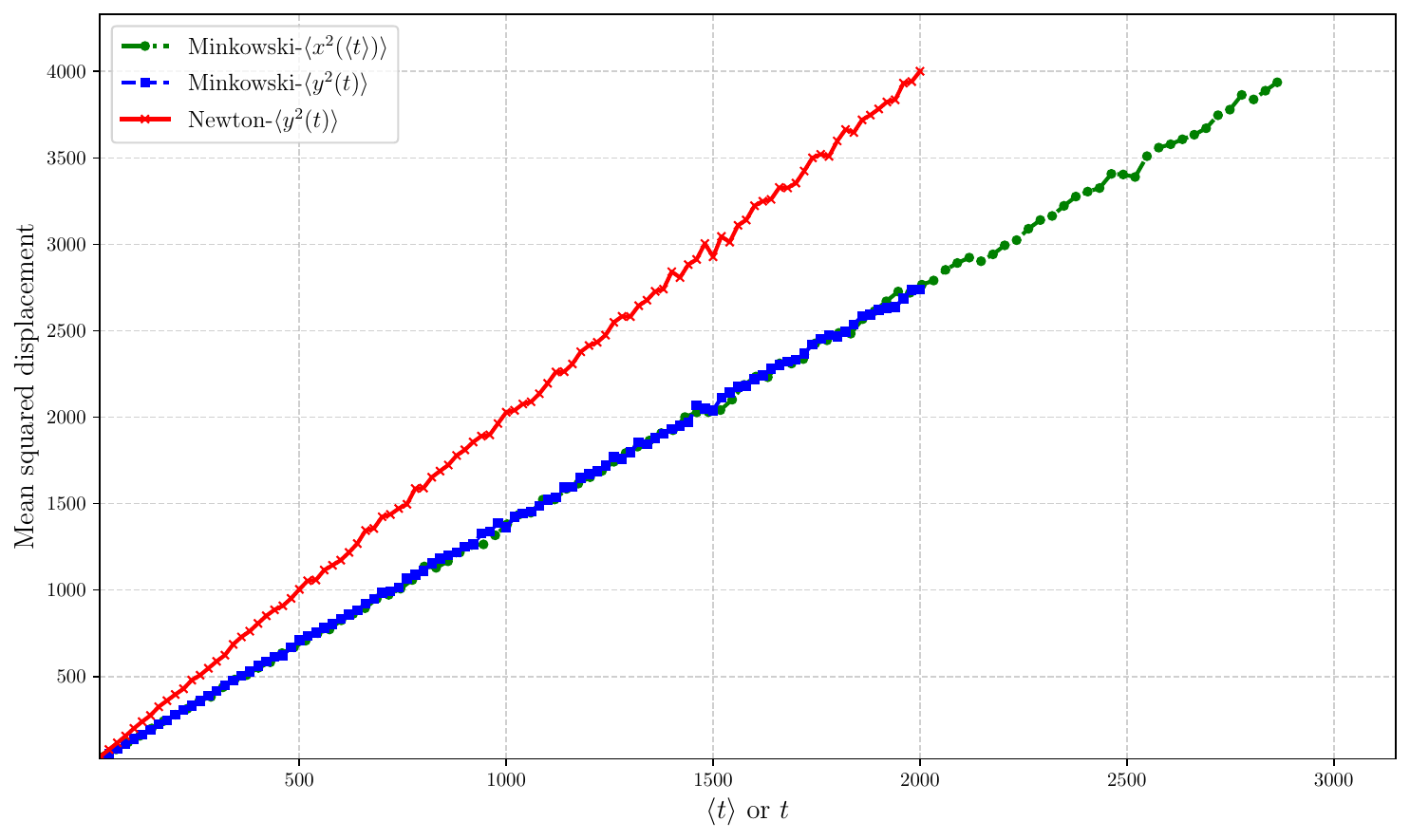}
\caption{Mean squared displacement over $\la t\ra$ and $t$}
\label{fig3}
\end{figure}
The final figure, Fig.~\ref{fig3}, presents the mean squared displacement with respect to $\la t\ra$. As previously mentioned, data obtained directly from simulations of LE$_{\tau}$ do not correspond to any experimentally observable quantities. However, such data can still be used to indirectly reconstruct observables. By combining $\langle x^{2}(\tau) \rangle$ and $\langle t(\tau) \rangle$ to eliminate the parameter $\tau$, the resulting curve aligns well with the dataset labeled by Minkowski-$\la y^{2}(t)\ra$. This indirectly confirms the validity of our approach, as starting the analysis with LE$_{\tau}$ for convenience and ultimately translating the results into the LE$_{t}$ proves to be well justified.

\section{Elimination of Fast Variables in Equation of Distribution}
\label{sec4}

While the previous analysis on the equations of motion provided a general approach to deriving effective dynamics across timescales, the statistical physics picture underlying the method remains less transparent. In this section, the focus is shifted to the level of the distribution function, and a same coarse-grained result is derived. A key advantage of this formulation is that it allows for a clearer identification of how and where the coarse-graining (i.e., adiabatic elimination) affects the covariance of the resulting equation.

\subsection{Kramers' trick}

Numerical simulations confirm that, under certain parameter settings, momentum acts as a fast variable relative to displacement, suggesting that Kramers' trick may be an appropriate tool in this context\cite{kramers1940brownian}. Kramers' trick provides a general procedure to project the full phase space dynamics onto position space, but its applicability to nonlinear Brownian motion is limited. The accuracy of the coarse-grained result depends crucially on the structure of the system; only in case the noise term of the underlying SDE is an additive noise,  does the method yield reliable results.

Fortunately, the LE$_{\tau}$ with multiplicative noise in Cartesian coordinates \eqref{langtau} transforms into a SDE with additive noise \eqref{langtau-theta} when expressed in terms of rapidity coordinates. Moreover, since $p^{1'}=m\sinh\vartheta$, the monotonic relationship ensures that equilibrium in momentum implies equilibrium in $\vartheta$ as well. Then, following the spirit of Kramers' adiabatic elimination and guided by dimensional analysis, one may reasonably conclude that the particle moves approximately along the straight line
\begin{align}\label{adiabatic-elim}
	x+\dfrac{m\vartheta}{\kappa}=\hat{x},
\end{align}
when $m/\kappa$ is sufficiently small and $\vartheta$ rapidly relaxes to its equilibrium value. If Eq.~\eqref{adiabatic-elim} is treated as a coordinate transformation, then
\begin{align}\label{coor-trans}
	\left\{\begin{aligned}
		x&=\hat{x}-\dfrac{m\hat{\vartheta}}{\kappa},\\
		\vartheta&=\hat{\vartheta},
	\end{aligned}\right.\qquad\Longrightarrow\qquad
	\left\{\begin{aligned}
		\dfrac{\partial}{\partial \hat{x}}&=\dfrac{\partial}{\partial x},\\
		\dfrac{\partial}{\partial \hat{\vartheta}}&=-\dfrac{m}{\kappa}\dfrac{\partial}{\partial x}+\dfrac{\partial}{\partial \vartheta}.
	\end{aligned}\right.
\end{align}
Therefore, $\(\frac{\partial}{\partial \vartheta}-\frac{m}{\kappa}\frac{\partial}{\partial x}\)$ is the tangent vector of the curves defined by new coordinate $\hat{x}=\text{const}$. The reduced Fokker-Planck equation \eqref{fp-theta} can be reformulated as
\begin{align}
	\dfrac{p^{\mu}}{m}\dfrac{\partial\varphi}{\partial x^{\mu}}
	=\(\dfrac{\partial}{\partial\vartheta}-\dfrac{m}{\kappa}\dfrac{\partial}{\partial x}\)&\(\dfrac{\kappa}{m}\sinh\vartheta \varphi+\dfrac{D}{2m^{2}}\dfrac{\partial\varphi}{\partial\vartheta}+\dfrac{D}{2m\kappa}\dfrac{\partial\varphi}{\partial x}\)\notag\\
	&+\dfrac{m}{\kappa}\dfrac{\partial}{\partial x}\(\dfrac{D}{2m\kappa}\dfrac{\partial\varphi}{\partial x}+\dfrac{\kappa}{m}\sinh\vartheta \varphi\),\label{re-fp}
\end{align}
allowing its structure to align more naturally with the new coordinate system. By performing the integration of both sides over the invariant volume element on the mass shell $(\Gamma_{m}^{+})_{\hat{x}}$, and invoking the adiabatic approximation, the expression becomes significantly simplified. Let us now examine each term individually.

The integration of the term on the left-hand side of Eq.~\eqref{re-fp} can be evaluated using the moving frame $\{e_{\alpha'}{}^{\mu}\}$,
\begin{align}\label{int-left}
	\int\eta_{(\Gamma_{m}^{+})_{\hat{x}}}\dfrac{p^{\mu}}{m}\dfrac{\partial\varphi}{\partial x^{\mu}}=\int\D\hat{\vartheta}\dfrac{p^{\alpha'}e_{\alpha'}{}^{\mu}}{m}\dfrac{\partial\varphi}{\partial x^{\mu}}=\int\D\hat{\vartheta}\cosh\hat{\vartheta}\ U^{\mu}\dfrac{\partial\varphi}{\partial x^{\mu}},
\end{align}
where the first equality follows from the fact that the volume element of the phase space satisfies $\D x\wedge\D p/|p_{0}|=\D x\wedge\D\vartheta=\D\hat{x}\wedge\D\hat{\vartheta}$, implying that the volume element on the mass shell is $\D\hat{\vartheta}$. In addition, $p^{\mu}$ is expressed in the moving frame. In the second equality, we substitute the components of $p^{\alpha'}$, use $e_{0'}{}^{\mu}=U^{\mu}$ and $\vartheta=\hat{\vartheta}$, and apply the fact that the integral of an odd function over the entire real axis vanishes.

Under the adiabatic approximation, the one particle distribution function admits an approximate separation of variables 
\begin{align}\label{coarse-grained-solution-phase-space}
	\varphi(t,x,p)\approx\rho(t,x)\phi(p)=\rho(t,x)\dfrac{\e^{\beta U_{\mu}p^{\mu}}}{2\mathsf{J}_{00}(\zeta)}=\rho(t,x)\dfrac{\e^{-\zeta\cosh\vartheta}}{2\mathsf{J}_{00}(\zeta)}.
\end{align}
It should be emphasized that the variable separation employed here holds only approximately. Treating it as exact and substituting it into the original equation would lead to an invalid identity. Substituting the adiabatic approximation into Eq.~\eqref{int-left} yields
\begin{align}
	\int\eta_{(\Gamma_{m}^{+})_{\hat{x}}}\dfrac{p^{\mu}}{m}\dfrac{\partial\varphi}{\partial x^{\mu}}\approx \dfrac{\mathsf{J}_{01}(\zeta)}{\mathsf{J}_{00}(\zeta)}U^{\mu}\partial_{\mu}\rho.
\end{align}

\begin{figure}[htbp]
\centering
\includegraphics[width=0.5\textwidth]{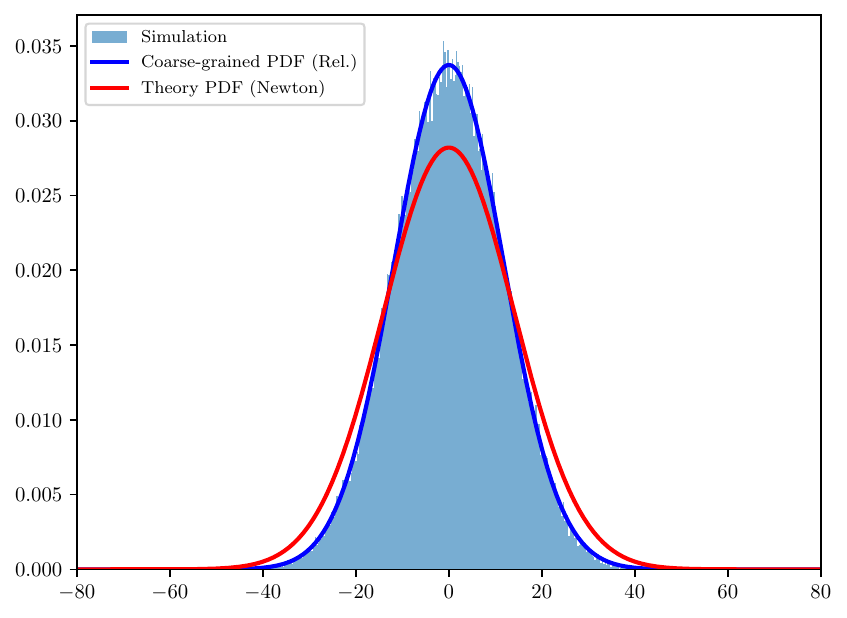}
\caption{Comparison between theoretical curve and simulation data at $t=100$.}
\label{fig4}
\end{figure}

The right hand side of Eq.~\eqref{re-fp} is much easier to evaluate. The integral of the first term yields
\begin{align}
	&\int\eta_{(\Gamma_{m}^{+})_{\hat{x}}}\(\dfrac{\partial}{\partial\vartheta}-\dfrac{m}{\kappa}\dfrac{\partial}{\partial x}\)\(\dfrac{\kappa}{m}\sinh\vartheta \varphi+\dfrac{D}{2m^{2}}\dfrac{\partial\varphi}{\partial\vartheta}+\dfrac{D}{2m\kappa}\dfrac{\partial\varphi}{\partial x}\)\notag\\
	=&\int\D\hat{\theta}\dfrac{\partial}{\partial\hat{\theta}}\(\dfrac{\kappa}{m}\sinh\vartheta \varphi+\dfrac{D}{2m^{2}}\dfrac{\partial\varphi}{\partial\vartheta}+\dfrac{D}{2m\kappa}\dfrac{\partial\varphi}{\partial x}\),
\end{align}
and since the distribution function vanishes as the rapidity tends to infinity, the boundary contribution is zero. Taking into account the adiabatic approximation, the integration of the second term is approximately given by
\begin{align}
	&\int\eta_{(\Gamma_{m}^{+})_{\hat{x}}}\dfrac{m}{\kappa}\dfrac{\partial}{\partial x}\(\dfrac{D}{2m\kappa}\dfrac{\partial\varphi}{\partial x}+\dfrac{\kappa}{m}\sinh\vartheta \varphi\)\notag\\
	\approx&\int\D\hat{\vartheta}\dfrac{m}{\kappa}\dfrac{\partial}{\partial \hat{x}}\(\dfrac{D}{2m\kappa}\dfrac{\partial\rho}{\partial \hat{x}}\dfrac{\e^{-\zeta\cosh\hat{\vartheta}}}{2\mathsf{J}_{00}(\zeta)}\)=\dfrac{D}{2\kappa^{2}}\dfrac{\partial^{2}\rho}{\partial \hat{x}^{2}}.
\end{align}
Putting together each term discussed above, we finally obtain the diffusion equation in spacetime derived via adiabatic elimination,
\begin{align}\label{coarse-grained-diffusion-equation}
	U^{\mu}\partial_{\mu}\rho=\dfrac{D}{2\kappa^{2}}\dfrac{\mathsf{J}_{00}(\zeta)}{\mathsf{J}_{01}(\zeta)}\dfrac{\partial^{2} \rho}{\partial \hat{x}^{2}}:=\text{D}_{\text{R}}\dfrac{\partial^{2} \rho}{\partial \hat{x}^{2}}.
\end{align}
The form of the final equation is strikingly similar to the Newtonian diffusion equation, 
\begin{align*}
	\dfrac{\partial\rho_{\text{N}}}{\partial t}=\dfrac{D}{2\kappa^{2}}\dfrac{\partial^{2} \rho_{\text{N}}}{\partial \hat{x}^{2}}:=\text{D}_{\text{N}}\dfrac{\partial^{2} \rho_{\text{N}}}{\partial \hat{x}^{2}}.
\end{align*}
With the present choice of observers and coordinate system, both equations admit Gaussian distribution as solution
\begin{align}
	\rho(t,\hat{x})=\dfrac{\exp\(-\frac{\hat{x}^{2}}{4\text{D}_{\text{R}}t}\)}{\sqrt{4\pi\text{D}_{\text{R}}t}},\qquad\rho_{\text{N}}(t,x)=\dfrac{\exp\(-\frac{x^{2}}{4\text{D}_{\text{N}}t}\)}{\sqrt{4\pi\text{D}_{\text{N}}t}}.
\end{align}
Fig.~\ref{fig4} clearly illustrates the relativistic correction to the diffusion behavior of a Brownian particle. Moreover, the simulation data obtained from LE$_{t}$, under the same parameter settings as those in Sec.~\ref{sec3.4}, can also be well described by the corresponding coarse-grained probability density function. Since the data were generated using LE$_{t}$, it is natural to question whether coarse-graining directly from the equation of $f_{t}(x,p)$ would yield more accurate results. The conclusion is that the equation obtained by applying the adiabatic approximation \eqref{adia-ft} to $f_{t}(x,p)$ is indeed consistent with Eq.~\eqref{coarse-grained-diffusion-equation}. Therefore, we do not repeat the derivation here. 

It is, of course, important not to be overly optimistic. Evidently, the coarse-graining procedure manifests a direct violation of the covariance inherent in the original equation of phase space distribution. However, from a purely coordinate transformation perspective, the adiabatic approximation appears not to have been introduced prior to Eq.~\eqref{coarse-grained-solution-phase-space}. This precisely illustrates the distinction between mathematics and physics: Coordinate transformation \eqref{coor-trans} does not hold over the entire phase space but is valid only within a limited region. However, when performing the integration, we extended it approximately to the whole phase space, which necessarily introduces a deviation from the original intended description of particle displacement behavior. Thus, although the curve in Fig.~\ref{fig4} provides a good fit to the simulation data, it does not represent the exact theoretical distribution. Moreover, the appearance of $\hat{x}$ as the argument of $\rho(t, \hat{x})$ further emphasizes that the distribution is only an approximation and does not directly describe the displacement $x$.

A final noteworthy observation is that the left-hand side of Eq.~\eqref{coarse-grained-diffusion-equation} maintains covariant structure. The breaking of covariance actually originates from the right-hand side. The well-behaved form on the left emerges naturally when expressing the Fokker-Planck equation with its left-hand side arranged as the action of a Liouville vector on the distribution function. Therefore, to obtain a covariant diffusion equation in spacetime, one possible approach would be to seek appropriate new coordinate curves whose tangent vectors remain covariant vectors in spacetime. This strategy may provide clues for future investigations.

\subsection{Characteristic parameters}

In the method of adiabatic elimination of fast variables, an important question arises: how fast must a variable be to qualify as being fast? In other words, what criteria determine the validity of the elimination procedure? In this subsection, we aim to address this question in the context of relativistic corrections to timescales.
%\begin{figure*}[hbt]
%\centering
%\includegraphics[width=0.8\textwidth]{./figure/juttner.pdf}
%\caption{J{\"u}ttner distribution with $\zeta=1$.}
%\end{figure*}

As discussed in Sec.~\ref{sec3.3}, the ratio $m/\kappa$ is commonly regarded as the characteristic timescale associated with the overdamped limit. However, this particular combination of system parameters fails to capture relativistic corrections. In fact, based on dimensional analysis, it is straightforward to construct a more general dimensionless quantity that better characterizes the relevant scale of the system. 

In any Brownian motion system to which adiabatic elimination can be applied, it is generally expected that two such quantities exist,
\begin{align*}
	p_{\sigma}:=\sqrt{\overline{(p-\overline{p})^{2}}},\qquad x_{\sigma}:=\sqrt{\la \(x(t)-\la x(t)\ra\)^{2}\ra},
\end{align*}
which denote the standard deviations of momentum and displacement, respectively. Since adiabatic elimination is applicable, $p_{\sigma}$ can be expected to remain time-independent. Based on $p_{\sigma}$ and $x_{\sigma}$, a reasonable dimensionless quantity characterizing the system's scale can then be constructed as follows
\begin{align*}
	\dfrac{p_{\sigma}}{\kappa x_{\sigma}}\sim 1.
\end{align*}
In the Newtonian case, 
\begin{align}
	\dfrac{p_{\sigma}}{\kappa x_{\sigma}}=\sqrt{\dfrac{m}{\kappa t}}\sim 1,
\end{align}
which clearly reproduces the conventional characteristic timescale. The advantage of this newly defined quantity lies in its dimensionless nature, allowing it to be freely used to characterize what can be considered ``fast'' or ``slow''.

It should be noted that for a general or unknown system, the quantities $x_{\sigma}$ and $p_{\sigma}$ cannot, in principle, be determined solely from the theoretical distribution. Instead, they must be extracted from experimental data. However, in the present relativistic case, the $x_{\sigma}$ and $p_{\sigma}$ calculated from the theoretical distribution offer a rough guideline for estimating the applicability of the method. Based on this, a theoretical expression can be constructed by substituting these quantities
\begin{align}\label{rel-criterion}
	\dfrac{p_{\sigma}}{\kappa x_{\sigma}}=\sqrt{\dfrac{\mathsf{J}_{21}(\zeta)}{\mathsf{J}_{00}(\zeta)}\dfrac{\zeta m}{\kappa t}}\sim 1.
\end{align}
According to the behavior of the special function $\mathsf{J}_{mn}(\zeta)$, the characteristic timescale in the relativistic case is actually larger than that in the Newtonian case.

\begin{figure}[hbt]
\centering
\includegraphics[width=\textwidth]{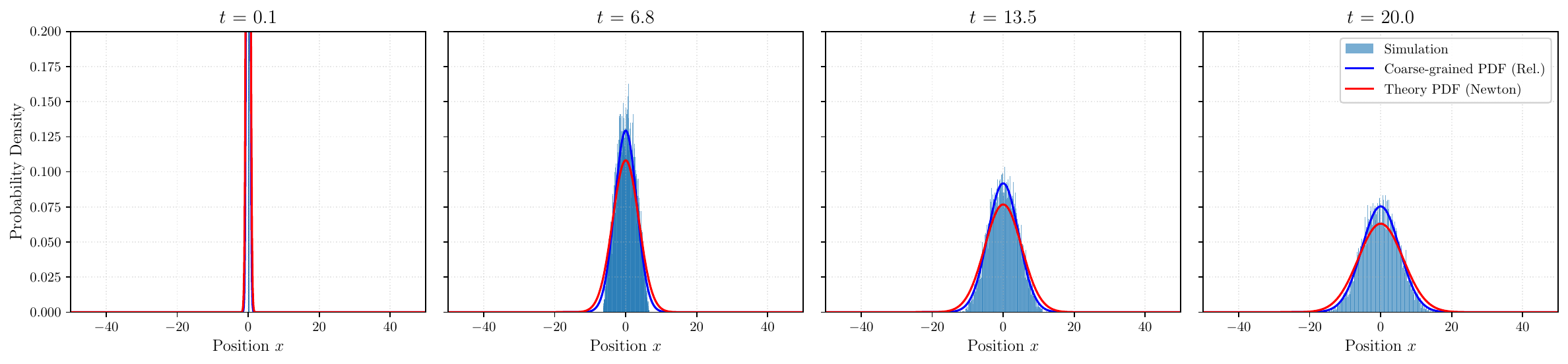}
\caption{Time evolution of the distribution function with $m=1, \kappa=1, \beta=1$.}
\label{fig5}
\end{figure}
It is more intuitive to illustrate this fact through numerical simulation. If we adopt the Newtonian criterion that the momentum variable is considered ``fast'' when $m/(\kappa t)<0.1$, the corresponding new criterion becomes $\frac{p_{\sigma}}{\kappa x_{\sigma}}<0.316228$. Under the parameter setting $m=1, \kappa=1, \beta=1$, the coarse-grained distribution is expected to match the simulation results around $t=10$ according to the Newtonian criterion, and it indeed provides a good fit to the simulation. However, as shown in Fig.~\ref{fig5}, a noticeable delay occurs, and the fitting only becomes accurate around $t=20$. This deviation is captured by the relativistic criterion above, which predicts a required time of $t\approx 38.59$. In fact, this suggests that the relativistic criterion might be overly conservative, since the distribution begins to match the simulation data earlier than this estimate, yet still notably later than what the Newtonian criterion would predict. This clearly indicates the inadequacy of the Newtonian criterion in the relativistic regime. A more relaxed and practical criterion, $\frac{p_{\sigma}}{\kappa x_{\sigma}}<0.4$, appears to be sufficient in capturing the onset of accurate coarse-graining.

It is worth noting that, although within this scale the coarse-grained result already agrees well with the actual behavior, it is still not the exact solution of the original equation in the full phase space. A more precise result requires an analysis similar to that of Chandrasekhar for the Klein–Kramers equation\cite{chandrasekhar1943stochastic}, which we leave for future work.

\subsection{Entropy of a relativistic Brownian particle}

Even in the non-relativistic context, the appropriate definition of entropy for a nonequilibrium system remains a subject of ongoing debate\cite{goldstein2019nonequilibrium}. In this work, we do not delve into the theoretical subtleties, but rather provide a comparative overview of several existing definitions.

The definition of entropy for a relativistic Brownian particle has been discussed \cite{dunkel2007relative,wang2024general}. If we adopt the definition proposed in \cite{wang2024general} for the present work, the resulting expression is given by
\begin{align}\label{entropy}
	S(t)&=-\int\(\D x\wedge\dfrac{\D p}{|p_{0}|}\)\dfrac{Z_{\mu}p^{\mu}}{m}\varphi(t,x,p)\log\varphi(t,x,p)\notag\\
	&\approx-\int\(\D x\wedge\D\vartheta\)\cosh\vartheta\rho(t,x)\phi(p)\log\[\rho(t,x)\phi(p)\]\notag\\
	&=-\dfrac{\mathsf{J}_{01}(\zeta)}{\mathsf{J}_{00}(\zeta)}\int\D x\rho(t,x)\log\rho(t,x)-\int\D\vartheta\cosh\vartheta\phi(p)\log\phi(p)\notag\\
	&=-\dfrac{\mathsf{J}_{01}(\zeta)}{\mathsf{J}_{00}(\zeta)}\int\D x\rho(t,x)\log\rho(t,x)-\dfrac{\mathsf{J}_{01}(\zeta)}{\mathsf{J}_{00}(\zeta)}\log \dfrac{\e^{-\zeta\mathsf{J}_{02}(\zeta)/\mathsf{J}_{01}(\zeta)}}{2\mathsf{J}_{00}(\zeta)},
\end{align}
where the second line incorporates the adiabatic approximation. The last term, which depends only on $\zeta$, consistently appears when the adiabatic approximation is applied, whether in the relativistic case or in  Newtonian case. This term violates the extensive property of entropy even for statistically independent particles. Since this term originates from the coarse-graining procedure, it should not play a significant role in the physical discussion. This may support the perspective presented in \cite{dunkel2007relative}, suggesting that relative entropy provides a more appropriate description of Brownian particles, particularly when considering entropy across different scales. The definition of relative entropy is also discussed in \cite{wang2024general}, where it can be verified that the last term is naturally eliminated.

After disregarding the last term, we can compare Eq.~\eqref{entropy} with the definition of Gibbs entropy in the Newtonian framework
\begin{align*}
	S_{\text{N}}(t)=-\int\D x \rho_{\text{N}}(t, x)\log\rho_{\text{N}}(t, x).
\end{align*}
Taking into account both the coefficient $\mathsf{J}_{01}(\zeta)/\mathsf{J}_{00}(\zeta)$ and the relativistic correction to the distribution function, it becomes straightforward to determine whether the resulting entropy is increased or decreased by relativistic effects. However, as $\zeta$ approaches infinity, the influence of relativistic corrections diminishes progressively and becomes negligible. Here it becomes clear that, in the context of thermodynamic systems, the so-called Newtonian limit involves not only the requirement that the motion of the system is much slower than the speed of light, but also that the coldness $\zeta$ of the system tends to infinity in order to recover the corresponding results.

\section{Coarse-Graining via Path Integrals}
\label{sec5}

Path integral representation based on the equation of motion for Brownian particles provides an alternative and effective approach to their statistical behavior. This section serves as a supplement to the main analysis of this paper, offering a more general framework that remains applicable even when the adiabatic elimination method breaks down. The general path integral formalism for relativistic Brownian motion has been established in the context of the generally covariant fluctuation theorem\cite{cai2025general,cai2025fluctuation}. Here, we focus on more specific aspects and therefore substitute the (1+1)-dimensional Minkowski spacetime setting discussed in this paper into the general expressions.

In the preceding subsections, geometric units were employed. However, since this subsection involves Taylor expansions of $k/mc$, we restore the speed of light $c$ in the expressions. This makes the structure of the expansion more transparent and facilitates the computation.

For practical calculations, Ito calculus is often more convenient than the Stratonovich formulation, and since there exists a definite relationship between the two, one can be transformed into the other. Accordingly, we reformulate the LE$_{t}$ \eqref{langt} in the Ito convention and present its discrete representation
\begin{subequations}\label{langt-discrete}
\begin{align}
\tilde{y}_{n+1}-\tilde{y}_{n}&=\dfrac{c^{2}\tilde{k}_{n}}{E_{n}}\D t,\\
\tilde{k}_{n+1}-\tilde{k}_{n}&=\sqrt{\dfrac{DE_{n}}{mc^{2}}}\circ_{I}\D\tilde{W}_{n}+\dfrac{D\tilde{k}_{n}}{2mE_{n}}\D t-\dfrac{\kappa \tilde{k}_{n}}{m}\D t,
\end{align}	
\end{subequations}
where $n$ denotes the discrete time step, and the increment $\D \tilde{W}_{n}$ follows the Gaussian distribution
\begin{align}
	\Pr[\D \tilde{W}_{n}=\D W_{n}]&=\dfrac{1}{(2\pi\D t)^{1/2}}\exp\[-\dfrac{(\D W_{n})^{2}}{2\D t}\].
\end{align}

Our ultimate goal is to construct the probability of the entire path $\Pr[\tilde{Y}_{[N]}=Y_{[N]}]$, where $Y=(y,k)$ is introduced to simplify the notation, and $N$ is the total number of time steps over the interval of interest. Owing to the Markov property, it can be decomposed as
\begin{align}\label{path-prob}
	\Pr[\tilde{Y}_{[N]}=Y_{[N]}]&=\Pr[\tilde{Y}_{0}=Y_{0}]\prod_{n=0}^{N-1}\Pr[\tilde{Y}_{n+1}=Y_{n+1}|\tilde{Y}_{n}=Y_{n}],
\end{align}
where $\Pr[\tilde{Y}_{0}=Y_{0}]$ denotes the initial probability density of the Brownian particle in phase space. To proceed, we need to transform the probability measure of $\D \tilde{W}_{n}$ into the conditional probability from step $n$ to step $n+1$.

According to the property of conditional probability and the discrete form of LE$_{t}$ \eqref{langt-discrete}, we have
\begin{align}
	&\Pr[\tilde{Y}_{n+1}=Y_{n+1}|\tilde{Y}_{n}=Y_{n}]=\Pr[\tilde{y}_{n+1}=y_{n+1},\tilde{k}_{n+1}=k_{n+1}|\tilde{Y}_{n}=Y_{n}]\notag\\
	=&\Pr[\tilde{y}_{n+1}=y_{n+1}|\tilde{k}_{n+1}=k_{n+1},\tilde{Y}_{n}=Y_{n}]\Pr[\tilde{k}_{n+1}=k_{n+1}|\tilde{Y}_{n}=Y_{n}]\notag\\
	=&\delta\(\dfrac{y_{n+1}}{y_{n}+c^{2}k_{n}\D t/E_{n}}-1\)\Pr[\tilde{k}_{n+1}=k_{n+1}|\tilde{Y}_{n}=Y_{n}].
\end{align}
Subsequently, to evaluate the second factor, the Radon–Nikodym theorem is applied to transform the probability measure across variables, yielding
\begin{align}
	\dfrac{1}{|(k_{0})_{n+1}|}\Pr[\tilde{k}_{n+1}=k_{n+1}|\tilde{Y}_{n}=Y_{n}]=\det\[\dfrac{\partial\D W}{\partial k}\](Y_{n+1},Y_{n})\Pr[\D \tilde{W}_{n}=\D W_{n}],
\end{align}
where the Jacobian determinant can be evaluated from the transformation
\begin{align*}
	\D W(\tilde{Y}_{n+1},\tilde{Y}_{n})=\D t\sqrt{\dfrac{mc^{2}}{DE_{n}}}\(\dfrac{\tilde{k}_{n+1}-\tilde{k}_{n}}{\D t}-\dfrac{D\tilde{k}_{n}}{2mE_{n}}+\dfrac{\kappa\tilde{k}_{n}}{m}\),
\end{align*}
and partial derivatives in the Jacobian determinant are taken with respect to the first argument of $\D W$. Substitution then leads to the explicit expression of the conditional probability,
\begin{align}\label{condi-prob}
	&\Pr[\tilde{k}_{n+1}=k_{n+1}|\tilde{Y}_{n}=Y_{n}]=\dfrac{E_{n+1}}{(2\pi\D t)^{1/2}}\sqrt{\dfrac{mc^{2}}{DE_{n}}}\exp\[-\dfrac{mc^{2}\D t}{2DE_{n}}\(\dfrac{\tilde{k}_{n+1}-\tilde{k}_{n}}{\D t}-\dfrac{D\tilde{k}_{n}}{2mE_{n}}+\dfrac{\kappa \tilde{k}_{n}}{m}\)^{2}\].
\end{align} 
With all terms in Eq.~\eqref{path-prob} explicitly determined, the full path probability is established. This expression enables the systematic computation of all statistical moments through integration.

\begin{figure}[hbt]
\centering
\includegraphics[width=0.6\textwidth]{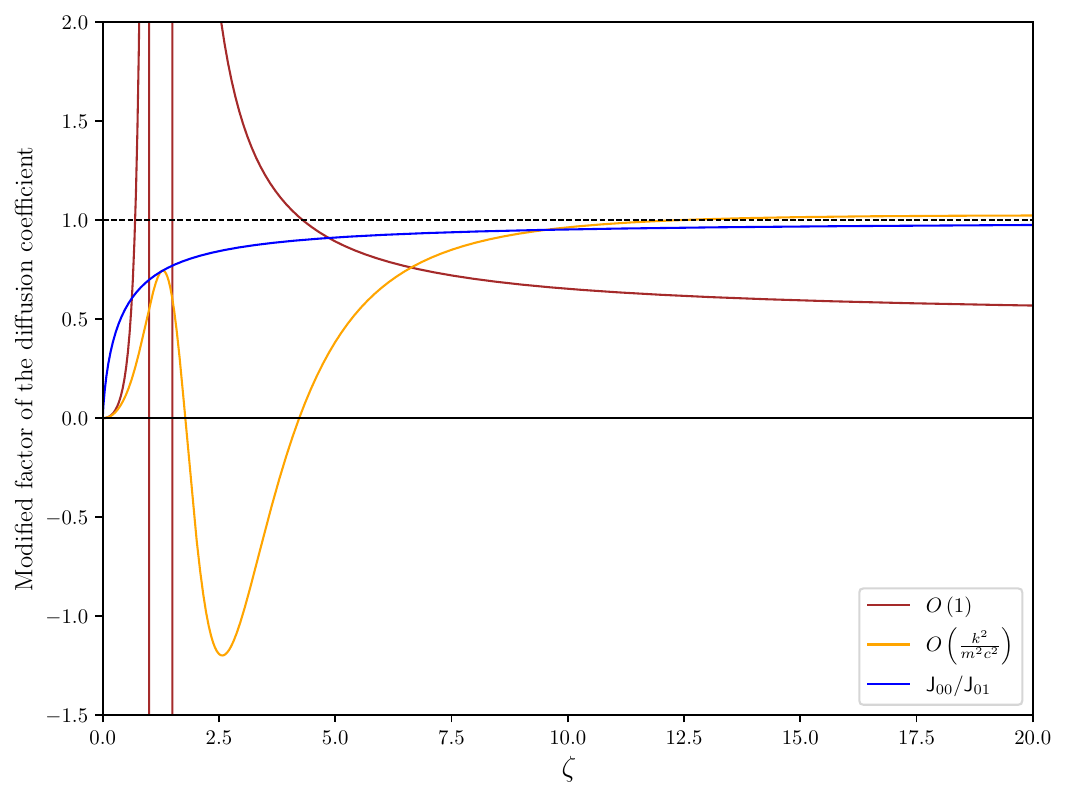}
\caption{Path integral approach versus adiabatic elimination method.}
\label{fig6}
\end{figure}

When we employ path integrals to directly compute correlation behavior in configuration space without considering momentum averaging, this approach is referred to as path integral coarse-graining. The key physical quantity in this framework, mean squared displacement, is defined as
\begin{align}\label{y-squared}
	\la y^{2}(t)\ra=\lim_{N\rightarrow\infty}\la y_{N}^{2}\ra=\lim_{N\rightarrow\infty}\int\(\prod_{n=0}^{N}\dfrac{\D y_{n}\D k_{n}}{|(k_{0})_{n}|}\)y_{N}^{2}\Pr[\tilde{Y}_{[N]}=Y_{[N]}],
\end{align}
where $t$ represents the time under consideration and $N$ denotes the number of temporal divisions. The probability density of path in this equation has been previously derived, thus in principle this computation can be performed directly. {However, practical calculations face technical complications. For example, if Eq.~\eqref{condi-prob} is substituted into Eq.~\eqref{y-squared} and the integration over $k_{N}$ is performed first, the result of this initial integration contains $E_{N-1}=c\sqrt{m^{2}c^{2}+k_{N-1}^{2}}$ and $E_{N-1}^{-1}$. The Gaussian integral of $E=c\sqrt{m^{2}c^{2}+k^{2}}$ or $E^{-1}$ over the momentum variable $k$ does not yield a closed-form expression.} Therefore, we expand $E$ as a Taylor series. Although the calculation is lengthy, it proceeds in a straightforward manner, so we present only the final results. The term retained up to $O(1)$ gives result
\begin{align}
	\lim_{t\rightarrow\infty}\dfrac{\la y^{2}(t)\ra}{t}=\dfrac{D}{\kappa^{2}}\dfrac{\zeta^{2}}{2\zeta^{2}-5\zeta+3},
\end{align}
while the expansion up to $O\(\frac{k^{2}}{m^{2}c^{2}}\)$ yields result
\begin{align}
	\lim_{t\rightarrow\infty}\dfrac{\la y^{2}(t)\ra}{t}=\dfrac{D}{\kappa^{2}}\dfrac{\zeta^{2}(2\zeta^{2}-12\zeta+15)}{(\zeta^{2}-3\zeta+3)(2\zeta^{2}-8\zeta+15)}.
\end{align}
In Fig.~\ref{fig6}, we compare the results from the first two orders of the path integral with those obtained under the adiabatic approximation. {Since $D/\kappa^{2}$ is the diffusion coefficient in position space for the Newtonian framework and also serves as a common factor in the relativistic expressions, the plot displays the dimensionless coefficients of $D/\kappa^{2}$ for comparison. The horizontal line at 1.0 represents the Newtonian diffusion coefficient.}

While a path integral  formally achieves arbitrary precision through systematic order-by-order evaluation, this demands considerable computational resources. A practical approach yielding accurate results efficiently, with rapid convergence, necessitates a small value of $k/mc$. Compared to the adiabatic elimination method discussed earlier, the advantage of the path integral approach lies in its, in principle, arbitrarily high accuracy—as long as one is willing to carry the expansion to sufficiently high orders. However, as seen from the graphical results, the expanded expressions, being polynomial in form, may exhibit unphysical zeros in certain regions, which clearly contradicts the expected behavior of a diffusion process. This issue arises entirely due to the truncation of the expansion; that is, insufficient expansion can lead to inaccurate results in parts of the domain. In contrast, the elimination method presented earlier imposes no restrictions on the range of $\zeta$. As long as the timescale separation holds, it provides a reliable and convenient approximation across the entire $\zeta$-domain. Therefore, each method has its own strengths and limitations.

\section{Conclusive Remarks}
\label{sec6}

This paper applies the method of adiabatic elimination in a relativistic context. Consistent relativistic corrections are obtained at both the levels of the equation of motion and the distribution function. Under identical parameter settings and within the regime where the adiabatic approximation is valid, relativistic diffusion proceeds more slowly than in the Newtonian case, and the linear relation $\la x^2 \ra \sim t$ remains approximately valid. A new characteristic parameter is introduced to quantify the validity range of the approximation, indicating that the relevant timescale for relativistic adiabatic elimination exceeds that of the Newtonian framework. As a complementary approach, the path integral formalism is also presented. All analyses are supported by numerical simulations, facilitating future efforts to identify experimental signatures.

It should be emphasized that although relativistic corrections do exist, their effects become significant only when the parameter $\zeta$ is sufficiently small. In practice, such corrections may be observable only in systems like electrons in terrestrial Tokamak devices, where $\zeta$ typically ranges from 20 to 500\cite{wesson2011tokamaks}. Another context in which the correction becomes particularly prominent is the process of Big Bang nucleosynthesis ($\zeta\sim 1$ for electrons\cite{fields2014big}). While most analyses of BBN rely on particle physics frameworks, it is important to recognize that at the macroscopic level, the suppression of particle diffusion due to relativistic effects alters the mean free path of reaction participants and consequently influences the efficiency of nucleosynthesis. Incorporating relativistic corrections in such settings is therefore essential. Moreover, this study focuses exclusively on the case of Minkowski spacetime. In scenarios where gravitational effects cannot be neglected(e.g., BBN), the coldness parameter $\zeta$ alone does not fully characterize the system. Dimensional analysis suggests that the gravitational constant $G$, the mass $M_{\text{hr}}$ of the heat reservoir, its characteristic scale $r_{\text{hr}}$, and the coldness parameter $\zeta$ together determine the relevant physical behavior. It is expected that an appropriate combination of these parameters may define a characteristic scale at which the adiabatic approximation remains valid, though this requires further investigation.

A careful reader may notice that covariance plays a central role in previous studies\cite{cai2023relativistic1,cai2023relativistic2,wang2024general,cai2025general,cai2025fluctuation}, whereas in the present work it appears to be less emphasized. The primary reason, as mentioned earlier, is that the coarse-graining procedure employed here breaks the manifestly covariant structure. However, this does not imply that a covariant coarse-graining scheme is impossible. On the contrary, developing such a formulation remains a promising direction for future research, since the fundamental laws of physics in spacetime are inherently covariant. Investigating Kramers' trick in higher-dimensional contexts and pursuing a geometric formulation of adiabatic elimination may serve as an initial step toward this goal. These issues will be addressed in future work.

\section*{Acknowledgement}

This work is supported by National Natural Science Foundation (Grant No. 12075059). One of the authors, Tao Wang, gratefully acknowledges stimulating discussions with Professor Liu Zhao's team. Appreciation is also extended to Dr. Wangzheng Zhang for valuable insights regarding potential theoretical applications.

\section*{Declaration of Competing Interest}

The authors declare that they have no known competing financial interests or personal relationships that could have appeared to influence the work reported in this paper.

\providecommand{\href}[2]{#2}\begingroup
\footnotesize\itemsep=0pt
\providecommand{\eprint}[2][]{\href{http://arxiv.org/abs/#2}{arXiv:#2}}

% \bibliographystyle{refformat}
% \bibliography{bib.bib}

\begin{thebibliography}{10}

\bibitem{einstein1919times}
A.~Einstein, ``What is the theory of relativity?,''{\emph{London Times} (1919) Nov. 28}.

\bibitem{hakim1965covariant}
R.~Hakim, ``A covariant theory of relativistic {Brownian} motion {I.} local equilibrium,''{\emph{J. Math. Phys.} {\bfseries 6} (1965) 1482}.

\bibitem{dudley1966lorentz}
R.~M. Dudley, ``Lorentz-invariant {Markov} processes in relativistic phase space,''{\emph{Ark. Mat.} {\bfseries 6} (1966) 241}.

\bibitem{debbasch1997relativistic}
F.~Debbasch, K.~Mallick and J.~P. Rivet, ``Relativistic {Ornstein--Uhlenbeck} process,''{\emph{J. Stat. Phys.} {\bfseries 88} (1997) 945}.

\bibitem{dunkel2009relativistic}
J.~Dunkel and P.~H{\"a}nggi, ``Relativistic {Brownian} motion,''{\emph{Phys. Rep.} {\bfseries 471} (2009) 1}.

\bibitem{herrmann2010diffusion}
J.~Herrmann, ``Diffusion in the general theory of relativity,''{\emph{Phys. Rev. D} {\bfseries 82} (2010) 024026}.

\bibitem{smerlak2012diffusion}
M.~Smerlak, ``Diffusion in curved spacetimes,''{\emph{New J. Phys.} {\bfseries 14} (2012) 023019}.

\bibitem{kremer2014diffusion}
G.~M. Kremer, ``Diffusion of relativistic gas mixtures in gravitational fields,''{\emph{Physica A} {\bfseries 393} (2014) 76}.

\bibitem{dunkel2005theory1}
J.~Dunkel and P.~H{\"a}nggi, ``Theory of relativistic {Brownian} motion: the (1+1)-dimensional case,''{\emph{Phys. Rev. E} {\bfseries 71} (2005) 016124}.

\bibitem{cai2023relativistic1}
Y.~Cai, T.~Wang and L.~Zhao, ``Relativistic stochastic mechanics {I}: Langevin equation from observer's perspective,''{\emph{J. Stat. Phys.} {\bfseries 190} (2023) 193}.

\bibitem{he2023heavy}
M.~He, H.~van Hees and R.~Rapp, ``Heavy-quark diffusion in the quark--gluon plasma,''{\emph{Prog. Part. Nucl. Phys.} {\bfseries 130} (2023) 104020}.

\bibitem{albertini2025stochastic}
E.~Albertini, A.~Nasiri and E.~Panella, ``Stochastic dark matter: Covariant {Brownian} motion from {Planckian} discreteness,''{\emph{Phys. Rev. D} {\bfseries 111} (2025) 023514}.

\bibitem{giardino2024first}
S.~Giardino, \emph{First-order thermodynamics of modified gravity}, Ph.D. thesis, 2024.

\bibitem{smerlak2013einstein}
M.~Smerlak, ``Einstein$^2$: Brownian motion meets general relativity,'' in \emph{Analogue Gravity Phenomenology}, pp.~385--398, Springer, (2013).

\bibitem{gao2019relativistic}
J.~H.~Gao and Z.~T.~Liang, ``Relativistic quantum kinetic theory for massive fermions and spin effects,''{\emph{Phys. Rev. D} {\bfseries 100} (2019) 056021}.

\bibitem{everitt2025}
M.~J. Everitt, ``From {Mass-Shell} {Factorisation} to {Spin}: {An} {Attempt} at a {Matrix-Valued} {Liouville} {Framework} for {Relativistic} {Classical} and {Quantum} {Phase-Spacetime},'' \emph{arXiv:}2505.03551v4,(2025).

\bibitem{van1985elimination}
N.~G. Van~Kampen, ``Elimination of fast variables,''{\emph{Phys. Rep.} {\bfseries 124} (1985) 69}.

\bibitem{cai2023relativistic2}
Y.~Cai, T.~Wang and L.~Zhao, ``Relativistic stochastic mechanics {II}: Reduced {Fokker-Planck} equation in curved spacetime,''{\emph{J. Stat. Phys.} {\bfseries 190} (2023) 181}.

\bibitem{klimontovich1994nonlinear}
Y.~L. Klimontovich, ``Nonlinear Brownian motion,''{\emph{Phys-Usp} {\bfseries 37} (1994) 737}.

\bibitem{risken1996fokker}
H.~Risken, \emph{The Fokker-Planck Equation}. Springer, 1996, ISBN: \href{https://link.springer.com/chapter/10.1007/978-3-642-61544-3\_4}{9783642615443}.

\bibitem{sekimoto2010stochastic}
K.~Sekimoto, \emph{Stochastic energetics}, vol.~799. Springer, 2010, ISBN: \href{https://link.springer.com/book/10.1007/978-3-642-05411-2}{9783642054112}.

\bibitem{kramers1940brownian}
H.~A. Kramers, ``Brownian motion in a field of force and the diffusion model of chemical reactions,''{\emph{Physica} {\bfseries 7} (1940) 284}.

\bibitem{chandrasekhar1943stochastic}
S.~Chandrasekhar, ``Stochastic problems in physics and astronomy,''{\emph{Rev. Mod. Phys.} {\bfseries 15} (1943) 1}.

\bibitem{goldstein2019nonequilibrium}
S.~Goldstein, D.~A. Huse, J.~L. Lebowitz and P.~Sartori, ``On the nonequilibrium entropy of large and small systems,'' in \emph{Stochastic Dynamics Out of Equilibrium: Institut Henri Poincar{\'e}, Paris, France, 2017}, pp.~581--596, Springer, 2019.

\bibitem{dunkel2007relative}
J.~Dunkel, P.~Talkner and P.~H{\"a}nggi, ``Relative entropy, haar measures and relativistic canonical velocity distributions,''{\emph{New J. Phys.} {\bfseries 9} (2007) 144}.

\bibitem{wang2024general}
T.~Wang, Y.~Cai, L.~Cui and L.~Zhao, ``General relativistic stochastic thermodynamics,''{\emph{SciPost Phys. Core} {\bfseries 7} (2024) 082}.

\bibitem{cai2025general}
Y.~Cai, T.~Wang and L.~Zhao, ``General relativistic fluctuation theorems,''{\emph{Phys. Lett. B} {\bfseries 860} (2025) 139220}.

\bibitem{cai2025fluctuation}
Y.~Cai, T.~Wang and L.~Zhao, ``Fluctuation theorems in general relativistic stochastic thermodynamics,''{\emph{Phys. Rev. E} {\bfseries 111} (2025) 024102}.

\bibitem{wesson2011tokamaks}
J.~Wesson and D.~J. Campbell, \emph{Tokamaks}, vol.~149. Oxford university press, 2011.

\bibitem{fields2014big}
B.~D. Fields, P.~Molaro and S.~Sarkar, ``Big {Bang} nucleosynthesis,'' in \emph{Review of Particle Physics}, PDG, ed., (2023).

\end{thebibliography}
\endgroup

\end{document}